\documentclass[a4paper]{article}

\usepackage[english]{babel}
\usepackage[utf8x]{inputenc}
\usepackage[T1]{fontenc}

\usepackage[a4paper,top=2.54cm,bottom=2.54cm,left=2.54cm,right=2.54cm,marginparwidth=1.75cm]{geometry}

\usepackage{amsmath}
\usepackage{graphicx}
\usepackage[colorinlistoftodos]{todonotes}
\usepackage[colorlinks=true, allcolors=blue]{hyperref}
\usepackage{parskip}
\usepackage{booktabs}

\usepackage{setspace}

\usepackage{cite}

\usepackage{amsmath}

\usepackage{graphicx}

\usepackage{multirow}
\usepackage{float}
\usepackage{etoolbox}
\usepackage[font=large,labelfont=bf]{caption}
\usepackage{mwe}

\title{\textbf{Effect of Metal Doping on the Visible Light Absorption, Electronic Structure and Mechanical Properties of Toxic-Free CsGeCl\textsubscript{3} Metal Halide}}

\author{\normalsize Md. Zahidur Rahaman\textsuperscript{\textdagger}, A.K.M. Akther Hossain\thanks{Corresponding author. \small \textsuperscript{\textdagger}\textit{\href{mailto:zahidur.physics@gmail.com}{zahidur.physics@gmail.com}} (M. Z. Rahaman); \small\textsuperscript{\textdaggerdbl}\textit{\href{mailto: akmhossain@phy.buet.ac.bd}{akmhossain@phy.buet.ac.bd}} (A.K.M. Akther Hossain).} ~\textsuperscript{\textdaggerdbl}\\\vspace{.01in}
\\\small \textit{\textsuperscript{\textdagger, \textdaggerdbl}Department of Physics}\\\small \textit{Bangladesh University of Engineering and Technology, Dhaka-1000, Bangladesh}}

\date{\small (27 June, 2018)} 

\begin{document}
\maketitle
\begin{abstract}
\addcontentsline{toc}{section}{Abstract}
\normalsize
\singlespace{
Toxic-free metal halide perovskites have become forefront for
commercialization of the perovskite solar cells and optoelectronic
devices. In the present study, for the first time we show that
particular metal doping in CsGeCl\textsubscript{3} halide can
considerably enhance the absorbance both in the visible and ultraviolet
light energy range. By using DFT based first principles method Mn and Ni
is doped at the Ge-site of CsGeCl\textsubscript{3} halide. We
investigate the detailed structural, optical, electronic and mechanical
properties of all the doped compositions theoretically. The study of
optical properties exhibits that the absorption edge of both Ni and
Mn-doped CsGeCl\textsubscript{3} is shifted toward the low energy region
(red shift) relative to the pristine one. An additional peak is observed
for both doped profile in the visible light energy region. The study of
mechanical properties ensures that both the doped samples are
mechanically stable and ductile as the pristine CsGeCl\textsubscript{3}.
The study of electronic properties shows that the excitation of photoelectrons is easier due to the formation of intermediate states in
Mn-doped CsGeCl\textsubscript{3}. As a result Mn-doped
CsGeCl\textsubscript{3} exhibits higher absorptivity in the visible
region than the Ni-doped counterpart. A combinational analysis suggests
that CsGe\textsubscript{1-x}Mn\textsubscript{x}Cl\textsubscript{3} is
the best lead free candidate among the inorganic prsovskite materials
for solar cell and optoelectronic applications.

}

\end{abstract}
\clearpage
\section*{1. Introduction}
\addcontentsline{toc}{section}{Introduction}
\large 
\doublespacing

Inorganic metal halide perovskite materials have been a topic of great
interest in the recent years due to their unique properties and
applications. These metal halides exhibit great promise not only to be
used as solar cell material but also show outstanding performance in
optoelectronics. These inorganic metal halides possess remarkable
optoelectronic properties including high optical absorption, small
carrier effective mass, tunable band gap, point defect, high charge
carrier mobility and long charge diffusion \cite{1, 2}. The applications
of these semiconductors are not limited in the branch of photovoltaics
and optoelectronics; they have a number of applications in
photodetector, LED (Light Emitting Diode) and devices which are used for
solar to fuel energy conversion \cite{3, 4, 5, 6}. Moreover, these
semiconducting materials are inexpensive and abundant in earth. Due to
this reason, the use of these materials in solar cells is convenient and
economic than the Si-based photovoltaic technology (PV-Technology)
\cite{1}.

The well known formula of metal halide perovskites is
ABX\textsubscript{3} (where, A = a cation, B = a metal ion and X =
a halogen anion). Most of the materials those exhibit remarkable
performances in this family contain lead (Pb). Therefore, the major
concern for application of these materials in practice is the toxicity
of Pb. In ambient condition the lead based metal halide perovskites
decompose to PbI\textsubscript{2} which is harmful for environment
\cite{7, 8, 9}. Hence, significant theoretical and experimental studies have
been carried out in the recent years to search for novel toxic free
perovskites by replacing metal cation for lead. Recently Roknuzzaman
\emph{et al}. carried out detailed investigation on this group of
materials by using theoretical method to find a suitable lead free
candidate \cite{10}. They perform the simulation work on the structural,
optical, electronic and mechanical properties of CsBX\textsubscript{3}
(B = Ge, Sn and X = Cl, Br, I) semiconductors and compare their results
with the lead containing semiconductors CsPbX\textsubscript{3} (X = Cl,
Br, I). After a combinational analysis they suggest that Ge-based
CsGeI\textsubscript{3} metal halide is the best lead free inorganic
metal halide perovskite semiconductor for optoelectronic and solar cell
application. However, the study of mechanical properties shows that the
proposed sample is brittle in nature and hence inconvenient for real
world application.

In 2016, B. Erdinc \emph{et al}. studied the electronic, optical,
thermodynamic and lattice dynamical properties of
CsGeCl\textsubscript{3} semiconductor for both paraelectric and
ferroelectric phases \cite{11}. Other experimental and theoretical study
suggests that CsGeCl\textsubscript{3} is a wide band gap semiconductor
with experimental band gap value 3.67 eV \cite{12, 13, 14, 15}. Because of this large band gap value CsGeCl\textsubscript{3} metal halide is not perfect for solar cell application. However, suitable metal doping in
CsGeCl\textsubscript{3} can reduce the band gap to such extent which is
suitable for the absorption of visible light energy. Therefore, in the
present study we aim to dope different transition metals in the Ge-site
of CsGeCl\textsubscript{3} metal halide for extending its absorption
spectra all over the range of the solar spectrum. We have studied the
optical, electronic and mechanical properties of Ni and Mn-doped
CsGeCl\textsubscript{3} by using the Density Functional Theory (DFT)
based theoretical method to find out a better candidate in this family
for solar cell and optoelectronic applications than the proposed best
lead free candidate CsGeI\textsubscript{3}. Finally, a thorough
comparison among the key properties of the metal doped
CsGeCl\textsubscript{3} and pristine CsGeCl\textsubscript{3} with the
CsGeI\textsubscript{3} metal halide is presented and discussed in
details.

\section*{2. Theoretical Methodology }
\addcontentsline{toc}{section}{Theoretical Methodology }
\large 
\doublespacing

The theoretical calculations are performed by using the Density
Functional Theory (DFT) based plane wave pseudopotential approach
\cite{16}. All the calculations are implemented by using the Cambridge
Serial Total Energy Package (CASTEP) executed within the Material
Studio-7.0 \cite{17, 25}. In order to get the doping effect in pure cubic
CsGeCl\textsubscript{3} metal halide, 2 × 2 × 2 supercell is constructed
which contains 40 atoms. As a result, the new chemical formula of
CsGeCl\textsubscript{3} perovskite can be written as
CsGe\textsubscript{1-x}M\textsubscript{x}Cl\textsubscript{3} (x = 0.125
and M = Ni, Mn). Generalized Gradient Approximation (GGA) proposed by
Perdew, Burke and Ernzerhof (PBE) is used for evaluating the exchange
correlation energy \cite{18}. The wave function is expanded up to 350 and 500 eV plane wave cutoff energy for doped and pristine samples,
respectively. For ensuring the criteria of convergence for both the
electronic properties calculation and geometry optimization 3 × 3 × 3
k-points and 10 × 10 × 10 k-points have been used for doped and pure
sample, respectively. For describing the electron ion interaction
Vanderbilt type ultrasoft pseudopotential is used \cite{19}. BFGS
(Broyden-Fletcher-Goldfarb-Shanno) relaxation scheme is used for
optimizing the crystal structure \cite{20}. The unit cell and atomic
relaxations are performed as long as the residual forces are below 0.03
eV/Å.

Finite strain theory \cite{21} executed within the CASTEP module is used
for evaluating the elastic constants of pristine and doped
CsGeCl\textsubscript{3} metal halide. We set 0.003 as the maximum strain
amplitude. Voigt-Reuss-Hill (VRH) averaging scheme \cite{22} is used for
obtaining the polycrystalline mechanical parameters from the evaluated
\emph{C\textsubscript{ij}} . The polycrystalline elastic moduli are
calculated by using Equations (1)-(7) given elsewhere \cite{23}. The
optical properties are calculated by using the CASTEP tool based on the
standard DFT Kohn-Sham orbitals \cite{24}. A scissor operator of 2.673 eV is applied in the calculation in order to compensate for the gap between the theoretical value (0.997 eV) and experimental value (3.67 eV) of the pure CsGeCl\textsubscript{3} band gap.

\section*{3. Results and Discussion}
\addcontentsline{toc}{section}{Results and Discussion}

\subsection*{3.1. Structural Properties }
\addcontentsline{toc}{subsection}{Structural Properties }
\large 
\doublespacing

\begin{figure}[H]
\centering
\includegraphics[width=1\textwidth]{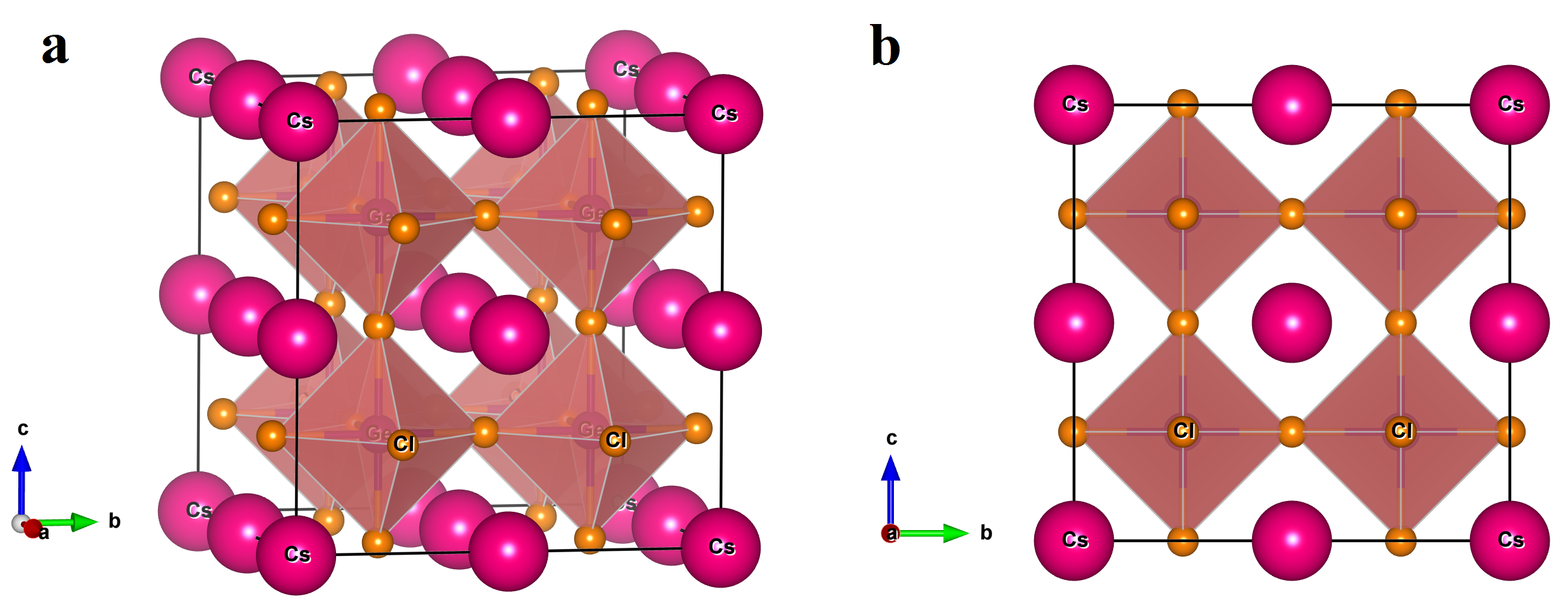}
\caption{The crystal structure (2 × 2 × 2 supercell) of CsGeCl\textsubscript{3} metal halide. (a) Three dimensional and (b) two dimensional view.} 
\label{fig:Fig. 1}
\end{figure}

Inorganic metal halide perovskite semiconductor CsGeCl\textsubscript{3}
belongs to the space group \emph{Pm3m} (221) with cubic crystal
structure \cite{10}. The unit cell consists of five atoms with only one
formula unit. The fractional coordinates of the Cs, Ge and Cl atoms are
(0, 0, 0), (0.5, 0.5, 0.5) and (0, 0.5, 0.5) with Wyckoff position
1\emph{a} , 1\emph{b} and 3\emph{c}, respectively. The experimental
lattice parameter and unit cell volume are listed in Table \ref{table 1} with the theoretical values calculated in this study. For inserting impurities in CsGeCl\textsubscript{3} perovskite a supercell having size 8 times the unit cell of the pristine sample is constructed as shown in Fig. \ref{fig:Fig. 1}. The supercell of CsGeCl\textsubscript{3} contains 40 atoms including 8 Cs atoms, 8 Ge atoms and 24 Cl atoms. The impurities are inserted in pure CsGeCl\textsubscript{3} by replacing one Ge atom by Ni/Mn atom (substitutional doping) which corresponds to the doping concentration of about 12.5 atom\%. The evaluated unit cell parameters of pristine and
doped samples are tabulated in Table \ref{table 1}.


\subsection*{3.2. Optical Properties}
\addcontentsline{toc}{subsection}{Optical Properties}
\large 
\doublespacing

In general, lead free (non-toxic) metal halide perovskites possess low reflectivity, high absorption coefficient and high optical conductivity than their lead containing counterpart. The study of optical properties is
essential for discovering a suitable material in this family for the
application in optoelectronic devices and solar cells. In this section
of the paper the important optical parameters including absorption
coefficient, reflectivity, real and imaginary part of dielectric
constant and optical conductivity of pristine and Ni/Mn-doped
CsGeCl\textsubscript{3} are analyzed and discussed in details.

\begin{figure}[H]
\centering
\includegraphics[width=1\textwidth]{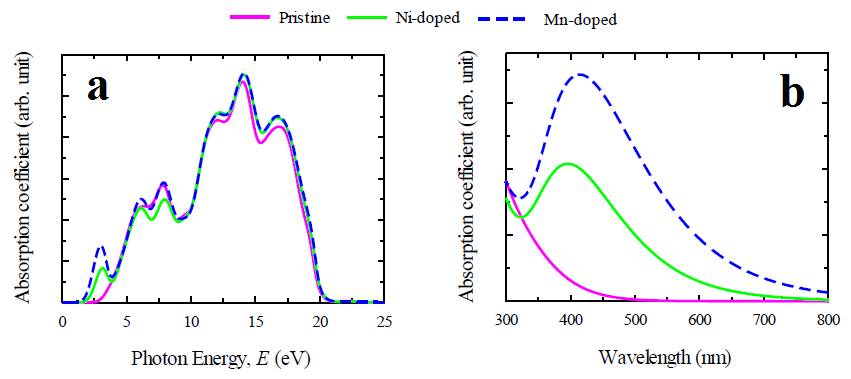}
\caption{Calculated light absorption spectra of pure and metal doped CsGeCl\textsubscript{3} perovskite. (a) Photon energy dependent and (b) wavelength dependent absorption coefficient.}  
\label{fig:Fig. 2}
\end{figure}

The evaluated absorption profiles of pristine and doped
CsGeCl\textsubscript{3} are illustrated in Fig. \ref{fig:Fig. 2}. The optical absorption coefficient is defined as the fraction of energy (wavelength) absorbed per unit length of the material. It also provides crucial information about the efficiency of optimum solar energy conversion of a material. Fig. \ref{fig:Fig. 2}(a) exhibits the photon energy dependent absorption coefficient of both pure and doped CsGeCl\textsubscript{3}. According to the result, the absorption edge of both Ni and Mn-doped CsGeCl\textsubscript{3} is shifted toward the low energy region (red shift) relative to the pristine one. An additional peak is observed for both doped profile in the low energy region. The absorption edge of Mn-doped CsGeCl\textsubscript{3} shift more toward the lower energy region than Ni-doped CsGeCl\textsubscript{3}. The pristine CsGeCl\textsubscript{3} shows no absorbance in the visible light region. The metal doping in pure CsGeCl\textsubscript{3} elevates the absorption coefficient to a great extent not only in the visible region but also in the ultraviolet region. For clear understanding the light absorption feature of CsGeCl\textsubscript{3} in the visible region, wavelength dependent absorption coefficient is presented in Fig. \ref{fig:Fig. 2}(b). As shown in Fig. \ref{fig:Fig. 2}(b) the Mn-doped CsGeCl\textsubscript{3} has widest absorption area than the Ni-doped CsGeCl\textsubscript{3}. The reason for the formation of broadest absorption area in Mn-doped sample will be discussed in details in the next section. In general, wide band gap
semiconductors can absorb ultraviolet light of solar spectrum which is
only 4\% of the total solar energy coming to the earth \cite{26}. The
visible light covers approximately 43\% of solar spectrum \cite{27}. Therefore, the intrinsic CsGeCl\textsubscript{3} (band gap = 3.67 eV) is
incapable to utilize the visible light energy for photovoltaic
conversion. Hence, prominent absorption in the visible region in
Mn-doped sample shows great promise for better utilization of solar
spectrum and may be increased the solar cell efficiency.

\begin{figure}[H]
\centering
\includegraphics[width=1\textwidth]{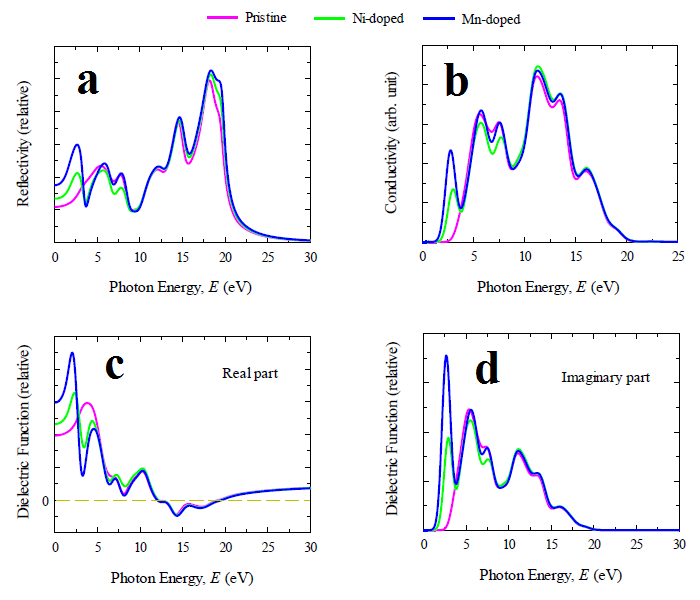} 
\caption{The simulated photon energy dependent (a) Reflectivity, (b) Conductivity, (c) Real part of dielectric function and (d) Imaginary part of dielectric function of pristine and transition metal doped CsGeCl\textsubscript{3} metal halide.} 
\label{fig:Fig. 3}
\end{figure}

Reflectivity is one of the vital optical properties of material for
photovoltaic application and is defined as the amount of light energy
reflected from the surface of a material with respect to the amount of
light energy incident on the surface of the material. The reflectivity
spectra of pristine and doped sample for photon energy up to 30 eV are
illustrated in Fig. \ref{fig:Fig. 3}(a). CsGeCl\textsubscript{3} shows low reflectivity in the whole energy range of solar spectrum. However, the reflectivity of all metal doped samples is nearly identical with the pristine one in the ultraviolet region. An additional peak is observed in the visible region for all doped samples whereas Mn-doped CsGeCl\textsubscript{3} has strong reflectivity in the low photon energy range than the Ni-doped sample. The dielectric function is characterized by the response of a material to incident light energy. The charge carrier recombination rate and hence the overall performance of optoelectronic devices depends upon the static value of dielectric function \cite{28}. A material with high dielectric constant has relatively less charge carrier recombination rate. As a result the overall performance of optoelectronic devices is enhanced. The real and imaginary part of evaluated dielectric function of pure and doped CsGeCl\textsubscript{3} is depicted in Figs. \ref{fig:Fig. 3}(c) and (d). It is evident that the Mn-doped sample shows relatively high dielectric constant than the pure and Ni-doped CsGeCl\textsubscript{3}. Therefore, in terms of dielectric constant Mn-doped CsGeCl\textsubscript{3} is a promising candidate for solar cell and optoelectronic application than the pristine one. Generally a material with higher band gap exhibit lower dielectric constants \cite{29}. Since metal doping in CsGeCl\textsubscript{3} decreases the band gap value (See electronic properties section), the metal doped samples show higher dielectric constant than the pure CsGeCl\textsubscript{3}. As shown in Figs. \ref{fig:Fig. 3}(c) and (d) overall dielectric profile (both real and imaginary part) of metal doped samples is almost identical in the high energy region (Ultraviolet zone) with pure sample. Additional peak of the real part of dielectric constant is observed in the visible light energy zone. The imaginary part of dielectric constant of all the samples goes to zero above 19 eV while the real part reaches approximately unity. This result implies
that both doped and pure CsGeCl\textsubscript{3} halide exhibit
transparency with slight absorption in the high energy zone (above 19
eV) (It is also evident from absorption coefficient graph {[}Fig.
\ref{fig:Fig. 2}(a){]}). Appearance of the sharp peak of the imaginary part of dielectric constant of metal doped sample in the visible region implies
the occurrence of strong absorption in this region \cite{30} which also
justifies the result obtained from absorption spectra of doped
CsGeCl\textsubscript{3} {[}Fig. \ref{fig:Fig. 2}{]}. Therefore, the investigation of the dielectric constant of pure and metal doped CsGeCl\textsubscript{3} suggests that both pristine and doped sample possess high transmissivity in the high energy region (above 19 eV) and metal doped samples possess nearly zero transmissivity in the visible region. This is the reason for high absorptivity of metal doped CsGeCl\textsubscript{3} (particularly Mn-doped CsGeCl\textsubscript{3}) in the visible region. However, the study of reflectivity spectra {[}Fig. \ref{fig:Fig. 3}(a){]} shows that Mn and Ni-doped CsGeCl\textsubscript{3} have slight higher reflectivity in the visible energy zone. Hence, further research should conduct to reduce the reflectivity of metal doped CsGeCl\textsubscript{3} in the visible region which may further increase the absorptivity as well as efficiency of solar cell. The optical conductivity is also defined as the photoconductivity. The conductivity spectra of doped and pristine CsGeCl\textsubscript{3} are illustrated in Fig. \ref{fig:Fig. 3}(b) up to 25 eV light energy. The optical conductivity of metal doped CsGeCl\textsubscript{3} is almost similar with the pure sample in the high energy region. A sharp peak is appeared in the visible light energy zone for both doped profile whereas Mn-doped sample exhibits large photoconductivity than Ni-doped sample. The appearance of the large photoconductivity in the visible region for metal doped CsGeCl\textsubscript{3} is a consequence of large absorptivity in the low energy region {[}Fig. \ref{fig:Fig. 2}{]}.


\subsection*{3.3. Electronic Properties}
\addcontentsline{toc}{subsection}{Electronic Properties}
\large 
\doublespacing

\begin{figure}[H]
\centering
\includegraphics[width=1\textwidth]{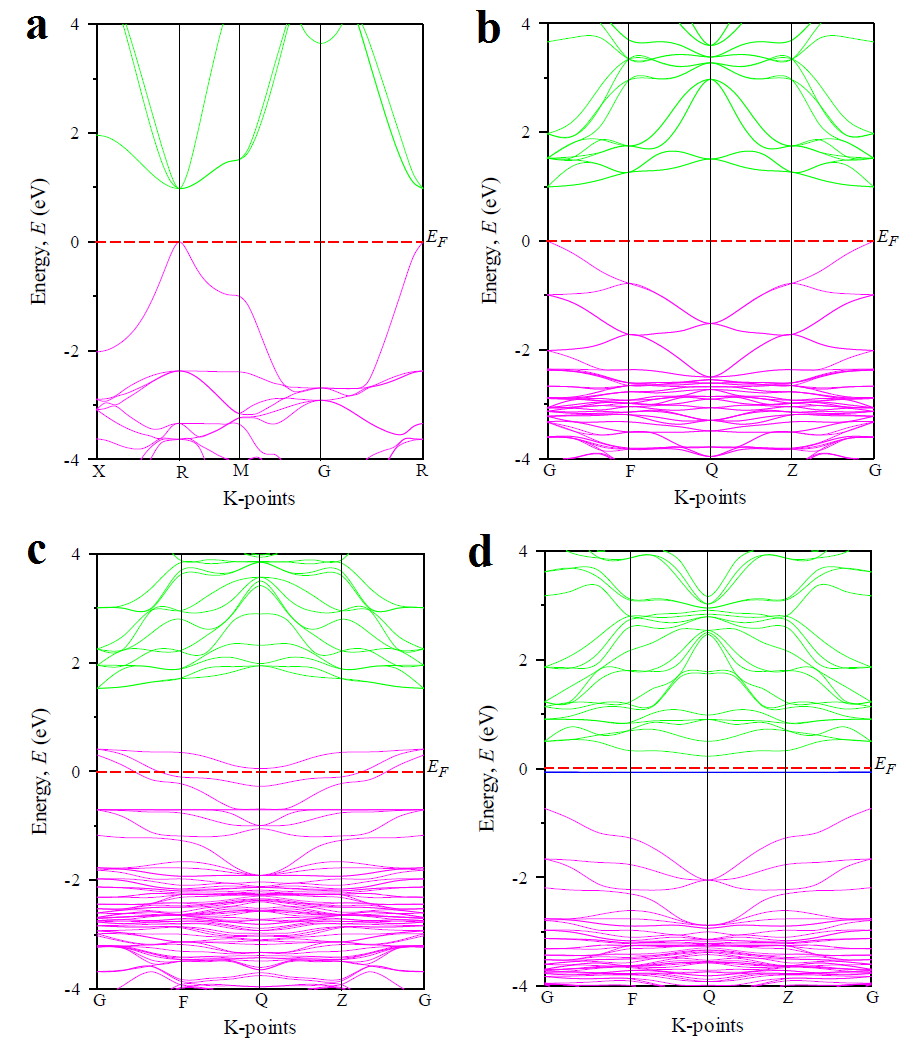} 
\caption{The band structure diagram of CsGeCl\textsubscript{3} metal halide calculated by using (a) pure single cell, (b) pure supercell, (c) Ni-doped and (d) Mn-doped sample.}
\label{fig:Fig. 4}
\end{figure}

For explaining the above optical features of metal doped
CsGeCl\textsubscript{3} halide the basic electronic properties including
band structure and density of states (DOS) of the studied samples are
calculated and discussed in this section. Fig. \ref{fig:Fig. 4} illustrates the electronic band structures of pristine and metal doped
CsGeCl\textsubscript{3} halide. The band structure diagram of pure
sample calculated using the single cell of CsGeCl\textsubscript{3} is
depicted in Fig. \ref{fig:Fig. 4}(a). As shown in the figure the conduction band minimum and valence band maximum lies at R (k-point), indicating the
direct band gap (0.97 eV) nature of pristine CsGeCl\textsubscript{3}.
This result exhibits good consistency with the evaluated value of 0.978
eV by Roknuzzaman \emph{et al}. \cite{10} implying the reliability of the present calculation. The band structure diagram illustrated in Fig. \ref{fig:Fig. 4}(b) is calculated by using the supercell (eight times) of
CsGeCl\textsubscript{3}. The computed band gap (direct) of pristine
CsGeCl\textsubscript{3} is 0.997 eV showing well agreement with the band
gap value calculated using single cell of CsGeCl\textsubscript{3}. It is
evident that the computed band gap value underestimates the
experimentally evaluated band gap value 3.67 eV \cite{15}. The reason can
be attributed to the well informed limitation of GGA. The Local Density Approximation (LDA) and LDA+U methods also underestimate the band gap value. Sometimes Heyd-Scuseria-Ernzerhof (HSE) hybrid potential may provide the band gap value close to the experimental one, but it is also not valid for all of the materials. Partial correction of the theoretical band gap value relative to the experimental one can be achieved by using GGA+U
approach. However, we only focus on the variation of band gap due to
different metal doping in CsGeCl\textsubscript{3} by ignoring the band
gap error of GGA approach. The band structure profile of metal doped
CsGeCl\textsubscript{3} halide is illustrated in Figs. \ref{fig:Fig. 4}(c) and \ref{fig:Fig. 4}(d), respectively. It is evident that impurity states emerge within the band gap of Ni-doped sample, whereas intermediate states appear within the band gap of Mn-doped CsGeCl\textsubscript{3}. In case of Ni-doped sample {[}Fig. \ref{fig:Fig. 4}(c){]}, vacant states are formed over the Fermi level. The valence band is extended into the higher energy region. The extension of valence band into the Fermi level can privilege the transition of electrons from valence band to conduction band. However, the gap between the maximum of the valence band and minimum of the conduction band is 1.10 eV which is larger than the band gap value of pristine CsGeCl\textsubscript{3} (0.997 eV). The enhancement of the band gap shows contradiction with the optical absorbance of Ni-doped
CsGeCl\textsubscript{3}. Here, the shift of Fermi level into the valence
band can be defined as the negative Burstein shift. This broadening of
the band gap can be ascribed to the phenomena know as Moss-Burstein
effect \cite{31}. Therefore, ignoring the Moss-Burstein shift the band
gap of Ni-doped sample is reduced and causes the occurrence of
absorbance in the visible region {[}Fig. \ref{fig:Fig. 2}(b){]}. As shown in Fig. \ref{fig:Fig. 4}(d) in case of Mn-doped CsGeCl\textsubscript{3} halide, an intermediate energy band is appeared in the band gap. The energy gap between the valence band and conduction band is 0.97 eV (similar with the pristine one) but the energy gap between the minimum of the conduction band and top of the intermediate states is 0.29 eV which is much smaller than the band gap value of pure CsGeCl\textsubscript{3}. Thus, the excitation of photo electrons is easier due to the formation of these intermediate states in Mn-doped CsGeCl\textsubscript{3}. As a result Mn-doped CsGeCl\textsubscript{3} exhibits higher absorptivity in the visible region than the Ni-doped counterpart. It should also note that the band gap of Mn-doped sample {[}Fig. \ref{fig:Fig. 4}(d){]} is indirect hence photons with this band gap energy can generate electron-hole pairs with the aid of phonons similar to silicon.

\begin{figure}[H]
\centering
\includegraphics[width=1\textwidth]{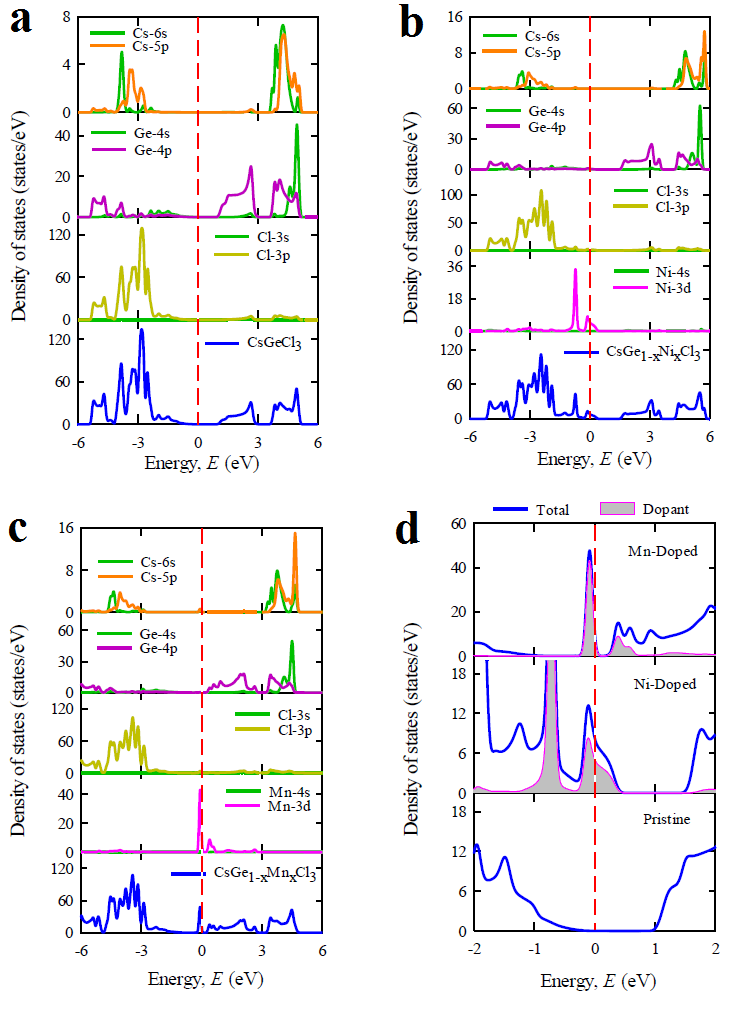}  
\caption{The total and partial density of states of CsGeCl\textsubscript{3} metal halide simulated by using (a) pure supercell, (b) Ni-doped sample, (c) Mn-doped sample and (d) dopant contribution at Fermi level.} 
\label{fig:Fig. 5}
\end{figure}

The total and partial density of states of pristine and doped
CsGeCl\textsubscript{3} is depicted in Fig. \ref{fig:Fig. 5}. Fig. \ref{fig:Fig. 5}(a) illustrates the DOS profile of pure sample. As shown in the figure the valence band is mostly composed of Cl-3p and Ge-4p orbitals with small contribution of Cs-6s and Cs-5p orbitals. The conduction band mainly consists of Ge-4p orbital with small contribution of Cs-6s and Cs-5p orbitals. After Ni doping in CsGeCl\textsubscript{3} a slight change is observed in the overall DOS profile as shown in Fig. \ref{fig:Fig. 5}(b). The composition of the valence band is almost similar with the pristine sample except an extra peak is appeared in the total DOS due to the formation of dopant states (Ni-3d) in the valence band of CsGeCl\textsubscript{3}. The composition of the conduction band is almost similar with the pristine one as no dopant peak is formed. Similar trend is observed for Mn-doped sample as shown in Fig. \ref{fig:Fig. 5}(c). The conduction band slightly shifts toward the lower energy region. Flat peak is observed in the conduction band due to the formation of dopant states (Mn-3d) in the conduction band. Fig. \ref{fig:Fig. 5}(d) illustrates the change of band gap due to the formation of new dopant states near Fermi level. In case of Ni-doped CsGeCl\textsubscript{3} the impurity energy states are formed over the Fermi level and mixed with the valence band maximum. These impurity states can trap the photoexcited holes which reduce the recombination rate of electrons and holes \cite{32}. In case of Mn-doped CsGeCl\textsubscript{3} the impurity energy states appear in the middle of the band gap. These intermediate states reduce the energy necessary for electron transition from valence band to conduction band. The valence electrons first excited to the impurity energy states (intermediate band) and then excited to the conduction band by consuming the visible light energy. These results explain the red shift of absorption spectra as shown in Fig. \ref{fig:Fig. 2}.


\subsection*{3.4. Mechanical Properties }
\addcontentsline{toc}{subsection}{Mechanical Properties }
\large 
\doublespacing

For ensuring the mechanical stability of metal doped
CsGeCl\textsubscript{3} halide the elastic constants of the doped phases
are calculated and discussed in this section in details. The calculated
elastic constants of pristine and doped CsGeCl\textsubscript{3} are
tabulated in Table \ref{table 2}. It is evident that all the compositions fulfill the well established Born stability criteria \cite{33} given as follows,

\emph{C\textsubscript{11}} \textgreater{} 0, \emph{C\textsubscript{44}} 
\textgreater{} 0, \emph{C\textsubscript{11}} --
\emph{C\textsubscript{12}} \textgreater{} 0 and
\emph{C\textsubscript{11}} + 2\emph{C\textsubscript{12}} \textgreater{}
0\\
Hence, both the doped phases are mechanically stable in nature. It is
also evident that the calculated elastic constants of pristine
CsGeCl\textsubscript{3} (by using supercell) agrees well with the
available theoretical result bearing the reliability of the present
investigation. The Cauchy pressure
(\emph{C\textsubscript{12}-C\textsubscript{44}} ) is a useful parameter
to predict the brittleness and ductility of materials. The negative
(positive) value of this parameter indicates the brittle (ductile)
nature of a compound \cite{30}. The computed values of the Cauchy
pressure of pristine and doped samples are positive {[}Table \ref{table 2}{]} implying that the doped samples are ductile as the pristine
CsGeCl\textsubscript{3} \cite{10}.

By using the computed elastic constants the most important mechanical
parameters of a compound such as the shear modulus \emph{G} , bulk
modulus \emph{B} , Young's modulus \emph{E} , \emph{B/G} ratio and
Poisson's ratio \emph{\(\nu\)} of pristine and doped samples are calculated
and listed in Table \ref{table 3}. It is also evident from Table \ref{table 3} that all the calculated mechanical parameters show good consistency with the other theoretically calculated results. The bulk modulus is one of the
essential mechanical parameter for getting idea about the stiffness of a
material. The calculated bulk modulus of all the composition is
comparatively low indicating the flexibility of all the samples. The
value of \emph{B} is slightly increased after Ni and Mn doping. However,
this lower value of \emph{B} ensures that it will be easier to make thin
film of Mn-doped CsGeCl\textsubscript{3} perovskite and hence suitable
for solar cell application. The shear modulus is used to get idea about
the plastic deformation of a material under external stress. As shown in
Table \ref{table 3} the value of G for all the samples is comparatively low. The value of G is slightly increased after Ni and Mn doping into
CsGeCl\textsubscript{3}. However, the lower value of shear modulus
implies that CsGe\textsubscript{1-x}Mn\textsubscript{x}Cl\textsubscript{3} is less rigid as the pristine one and hence can be drawn into desired shape. Similar trend is noticed for Young's modulus of all the compositions.

For explaining the bonding nature and plasticity of a material the
Poisson's ratio is another useful parameter. The calculated value of
\emph{\(\nu\)} for all the compositions is 0.26 implying the existence of
central force in all the samples \cite{34}. However, all the samples
should have predominant ionic feature as the value of \emph{\(\nu\)} is very close to the critical value 0.25 which indicates an ionic crystal. The
Poisson's ratio is another useful indicator of ductility and brittleness
of materials. The critical value for separating the ductility and
brittleness of a material is 0.26 \cite{35}. Surprisingly the computed
value of \emph{\(\nu\)} of all the samples is exactly 0.26 and hence implying the ductile nature of all the compositions. The ratio between bulk
modulus and shear modulus is usually known as Pugh's ratio which is also
used to predict the failure mode of a material. In this case, the
critical value for separating the ductility and brittleness of a
material is 1.75 \cite{36}. As shown in Table \ref{table 3} all the studied
compositions are ductile in nature as the value of \emph{B/G} is greater
than the critical value. The value of \emph{B/G} is decreased after Ni
and Mn doping in CsGeCl\textsubscript{3}. However, the value of Pugh's
ratio of metal doped CsGeCl\textsubscript{3} perovskite is still greater
than the critical value and hence should be exhibited the ductile manner
as the pristine one.


\subsection*{3.5. Environment-Friendly Perovskites  }
\addcontentsline{toc}{subsection}{Environment-Friendly Perovskites}
\large 
\doublespacing

Among the lead free inorganic perovskite materials suitable for solar
cell and optoelectronic applications CsGeI\textsubscript{3} perovskite
compound is proposed to be the best candidate \cite{10}. Though the band
gap of CsGeI\textsubscript{3} is suitable for the absorption of both
visible and ultraviolet light energy of the solar spectrum, it is
slightly brittle. Hence fabrication of solar cell by using
CsGeI\textsubscript{3} halide is difficult. On the other hand,
CsGeCl\textsubscript{3} is ductile but possesses large band gap which is
not suitable for the absorption of visible light energy of the solar
spectrum. Metal doping in CsGeCl\textsubscript{3} can solve the problem.

According to the present study, small portion of Ni doping in the Ge
site of CsGeCl\textsubscript{3} can elevate the absorption in the
visible region to a great extent. However, the Mn-doped
CsGeCl\textsubscript{3} exhibits very high absorption not only in the
visible region but also in the ultraviolet region. The study of
mechanical properties ensures that the proposed sample is ductile and
possesses low bulk modulus and hence possible to make thin film. It also
possesses high photoconductivity. The comparison among the key
properties of pure and metal doped CsGeCl\textsubscript{3} with
CsGeI\textsubscript{3} is presented in Table \ref{table 4}. From Table \ref{table 4} it is clear that CsGe\textsubscript{1-x}Mn\textsubscript{x}Cl\textsubscript{3} is
the best lead free candidate among inorganic prsovskite materials for
solar cell and optoelectronic applications.

\section*{4. Conclusions}
\addcontentsline{toc}{section}{Conclusions}
\large 
\doublespacing

In summary, by using the Density Functional Theory dependent
\emph{ab-initio} method the optical, electronic and mechanical
properties of metal doped CsGeCl\textsubscript{3} halide is investigated
and discussed in details. The intrinsic CsGeCl\textsubscript{3} is
incapable to utilize the visible light energy for photovoltaic
conversion due to its large band gap energy 3.67 eV. Particular metal
doping in CsGeCl\textsubscript{3} halide considerably enhance the
absorbance and photoconductivity both in the visible and ultraviolet
light energy range. However, due to the formation of intermediate energy
band Mn-doped CsGeCl\textsubscript{3} exhibits higher absorptivity and
photoconductivity in the visible region than the Ni-doped counterpart.
Hence, prominent absorption in the visible region of Mn-doped sample
shows great promise for better utilization of solar spectrum and may be
increased the solar cell efficiency. The study of mechanical properties
ensures that the Mn-doped CsGeCl\textsubscript{3} is ductile and
possesses low bulk modulus. Therefore, it is also possible to make thin
film by using
CsGe\textsubscript{1-x}Mn\textsubscript{x}Cl\textsubscript{3}. The
comparison among the key properties of pure and metal doped
CsGeCl\textsubscript{3} with the best known inorganic metal halide
perovskite CsGeI\textsubscript{3} suggests that
CsGe\textsubscript{1-x}Mn\textsubscript{x}Cl\textsubscript{3} is the
best lead free candidate among inorganic prsovskite materials for solar
cell and optoelectronic applications.

\section*{Acknowledgments} 
\addcontentsline{toc}{section}{Acknowledgments}
\large 
\doublespacing

This work is carried out in the Solid State Physics Research Laboratory, Department of Physics, Bangladesh University of Engineering and Technology, Dhaka, Bangladesh.

\begin{center}
\begin{tikzpicture}
\draw (-2,0) -- (2,0);
\filldraw [black,very thick] (0,0) (-1,0) -- (1,0);
\filldraw [black,ultra thick] (0,0) (-.5,0) -- (.5,0);
\end{tikzpicture}
\end{center}

\renewcommand{\section}[2]{}%

\clearpage

\clearpage

\begin{table*}[ht]
\doublespacing
\large
\centering 

\caption{\large The theoretical and experimental unit cell parameters
of pristine and doped CsGeCl\textsubscript{3} perovskite}
\label{table 1}
\bigskip

\begin{tabular}[]{@{}cccccc@{}}
\hline
Properties & & CsGeCl\textsubscript{3} & &
CsGe\textsubscript{1-x}Ni\textsubscript{x}Cl\textsubscript{3} &
CsGe\textsubscript{1-x}Mn\textsubscript{x}Cl\textsubscript{3}
\tabularnewline
\cmidrule(r){2-4}
& This study & Expt. \cite{15} & Calc. \cite{10} & &\tabularnewline
\hline
\emph{a\textsubscript{0}} (Å ) & 5.317 & 5.434 & 5.314 & 5.258 &
5.269\tabularnewline
\emph{V\textsubscript{0}} (Å\textsuperscript{3}) & 150.31 & 160.45 &
150.10 & 145.36 & 146.27\tabularnewline
\emph{B\textsubscript{0}} (GPa) & 29.69 & - & - & 33.60 &
36.96\tabularnewline
\hline
\end{tabular}

\end{table*}

\begin{table*}[ht]
\doublespacing
\large
\centering 

\caption{\large The evaluated elastic constants
\emph{C\textsubscript{ij}} (GPa) and Cauchy pressure of pristine and
doped CsGeCl\textsubscript{3} halide.}
\label{table 2}
\bigskip

\begin{tabular}[]{@{}ccccc@{}}
\hline
Phase & \emph{C\textsubscript{11}} & \emph{C\textsubscript{12}} &
\emph{C\textsubscript{44}} &
\emph{C\textsubscript{12}-C\textsubscript{44}}\tabularnewline
\hline
CsGeCl\textsubscript{3} & 55.02 & 13.14 & 11.73 & 1.41\tabularnewline
CsGe\textsubscript{1-x}Ni\textsubscript{x}Cl\textsubscript{3} & 54.16 &
14.60 & 13.36 & 1.24\tabularnewline
CsGe\textsubscript{1-x}Mn\textsubscript{x}Cl\textsubscript{3} & 56.68 &
14.36 & 13.33 & 1.03\tabularnewline
CsGeCl\textsubscript{3} \cite{10} & 54.93 & 13.08 & 11.99 &
-\tabularnewline
\hline
\end{tabular}

\end{table*}

\begin{table*}[ht]
\doublespacing
\large
\centering 

\caption{\large The evaluated mechanical parameters of pristine and
doped CsGeCl\textsubscript{3} halide.}
\label{table 3}
\bigskip

\begin{tabular}[]{@{}cccccc@{}}
\hline
Phase & \emph{B} (GPa) & \emph{G} (GPa) & \emph{E} (GPa) & \emph{\(\nu\)} &
\emph{B/G}\tabularnewline
\hline
CsGeCl\textsubscript{3} & 27.11 & 14.82 & 37.60 & 0.26 &
1.82\tabularnewline
CsGe\textsubscript{1-x}Ni\textsubscript{x}Cl\textsubscript{3} & 27.78 &
15.63 & 39.48 & 0.26 & 1.77\tabularnewline
CsGe\textsubscript{1-x}Mn\textsubscript{x}Cl\textsubscript{3} & 28.46 &
16.05 & 40.53 & 0.26 & 1.77\tabularnewline
CsGeCl\textsubscript{3} \cite{10} & 27.03 & 15.02 & 38.02 & 0.27 &
1.80\tabularnewline
\hline
\end{tabular}

\end{table*}

\begin{table*}[ht]
\doublespacing
\large
\centering 

\caption{\large Comparison among the key properties of pristine and
doped CsGeCl\textsubscript{3} with CsGeI\textsubscript{3}.}
\label{table 4}
\bigskip

\begin{tabular}[]{@{}ccccc@{}}
\hline
Properties & CsGeCl\textsubscript{3} & CsGeI\textsubscript{3} & CsGe\textsubscript{1-x}Ni\textsubscript{x}Cl\textsubscript{3} & CsGe\textsubscript{1-x}Mn\textsubscript{x}Cl\textsubscript{3} \tabularnewline
& \cite{10}, [This] & \cite{10} & &\tabularnewline 
\hline
& High in  & High in  & High in
 & High in \tabularnewline
Optical absorption & UV-region. & UV-region. & UV-region. & UV-region \tabularnewline
\cmidrule(r){2-5}
& Missing in  & Medium in  & Medium in
 & High in \tabularnewline
& visible region. & visible region. & visible region. & visible region \tabularnewline  
\hline
Photoconductivity & Medium & High & Medium & High\tabularnewline
Failure mode & Ductile & Brittle & Ductile & Ductile\tabularnewline
Suitability for solar cells & Not good & Good & Better &
Best\tabularnewline
\hline
\end{tabular}

\end{table*}

\end{document}